# On the Magnetism of C14 Nb(Fe$_{89.4}$Al$_{10.6}$)$_2$ Laves Phase Intermetallic Compound

Stanisław M. Dubiel[a*] and Israel Felner[b]

[a]AGH University of Science and Technology, Faculty of Physics and Applied Computer Science, al. A. Mickiewicza 30, 30-059 Kraków, Poland, [b]Racah Institute of Physics, The Hebrew University, Jerusalem, Israel 91904

## Abstract

C14 Nb(Fe$_{89.4}$Al$_{10.6}$)$_2$ Laves phase intermetallic compound was investigated by DC magnetization (M) measurements performed in the temperature (*T*) interval of 20 to 175 K, under an applied magnetic field (*H*) ranging between 50 and 1250 Oe. Magnetization curves were recorded in the field-cooled (FC) and in the zero-field-cooled (ZFC) modes. They clearly showed an irreversible character that vanished at *H*=1250 Oe. Both magnetization curves exhibited well-defined peaks around $T_N$ =72.3 K whose positions were *H*-independent, so they were identified as the compound's Néel temperature. The existence of irreversibility which decreases with *H* testify to a re-entrant character of magnetism in the studied compound. An increase of both M$_{FC}$ and M$_{ZFC}$ observed below $T_N$ likely indicates a mixed i.e. ferromagnetic and antiferromagnetic ground magnetic state of the studied system.

* Corresponding author: Stanislaw.Dubiel@fis.agh.edu.pl (S. M. Dubiel)

# 1. Introduction

Magnetic properties of the hexagonal C14 Laves phase $NbFe_2$ have been subject of numerous experimental [1-12] and theoretical [13-17] studies. The interest in the system follows from the fact that its magnetic state turned out to be very delicate and it can be easily impaired by external magnetic field $H > 6$ kOe or by slight deviation from the stoichiometry. Consequently, many different pictures of the $NbFe_2$ magnetism can be found throughout the literature. Initially, using magnetization and Mössbauer-effect measurements, Shiga and Nakamura concluded that the ground state of $NbFe_2$ is strongly enhanced Pauli paramagnet [1]. Yamada and Sakata reported that the stoichiometric compound had a weak antiferromagnetic (AF) ground state with a Néel temperature ($T_N$) of about 10 K. [2]. Later, based on magnetization and nuclear magnetic resonance results, Yamada et al. concluded, that the compound had the AF ground state but with AF and ferromagnetic (FM) fluctuations [3]. Crook and Cywinski investigated $Nb_{1-x}Fe_{2-x}$ samples ($-0.04 \leq x \leq 0.04$). They outlined a magnetic phase diagram in which the AF phase exists in the close vicinity of $x=0$ and $T_N$ ranges between ~10 K ($x \approx -0.012$) and ~27K ($x \approx 0.008$). Furthermore, they claim that the magnetic ground state is a mixture of AF and FM phases [4]. The most recent version of the magnetic phase diagram, [5] only partly agrees with the one reported in Ref. 4. Although in both phase diagrams the FM state exists for Fe- and Nb-over doped compounds, but the border $x$-values and those of the Curie temperature, $T_C$, are significantly different. The most important discrepancy occurs in the vicinity of $x=0$, where, following Crook and Cywinski, the AF and FM phases coexist, while according to the authors of Ref. 5 the two FM phases are separated by a paramagnetic phase with a quantum critical point at $x \approx -0.015$, and a spin-density waves phase exist for $-0.015 \leq x \leq 0.003$ at $T=0$ K.

Based on spin-dependent Compton scattering study supported by *ab initio* electronic structure calculations a ferri-magnetic ordering was recently suggested for non-stoichiometric $Nb_{0.985}Fe_{2.015}$ compound [6]. Noteworthy, some features typical of re-entrant spin-glasses (RSG) were observed both for the stoichiometric compound [7] as well as for the one with Fe-excess [8] Our current DC and AC magnetization susceptibility measurements on a $Nb_{0.975}Fe_{2.025}$ sample gave clear evidences in favor of the RSG behavior [9]. From all these studies it may be concluded that observed differences in the magnetic properties have its origin in the departure from stoichiometry.

Investigation of an effect of substitutional elements on the magnetic behavior of the $NbFe_2$ may be helpful in getting a deeper insight into its magnetic properties. Yamada et al. studied such effect in $Nb(Fe_{1-x}Al_x)_2$ samples [10] and found that the effect depends on the content of Al: for $x < 0.05$ the weak FM ordering occurs with Curie temperature ($T_C$) up to 5 K, while for $x = 0.05 – 0.4$ the ordering is antiferromagnetic with the maximum $T_N \approx 73$ K for $x \approx 0.2$. Yamada et al. also investigated the effect of 3d-elements (V, Cr, Mn, Co, Ni) and they concluded that substituting less than 1 at.% of V, Cr or Mn resulted in a weak FM state, while the same amount of Co and Ni created a non-magnetic ground state [11]. Paduani et al. studied the $Nb(Fe_{1-x}Co_x)_2$ system in the range $0 \leq x \leq 0.55$ and concluded that all three samples ($x$=0.10, 0.33 and 0.55) were weakly ferromagnetic with $T_C$ ranging between 14 and 35 K [12]. Furthermore, they found that the field-cooled (FC) and zero-field-cooled (ZFC) magnetization curves were irreversible indicating thereby a frustration of magnetic moments. This finding being at cross with the result reported in Ref. 11 challenges further studies on the effect of substitutional elements on magnetic properties of the $NbFe_2$ Laves phase. In this paper we report on dc magnetization measurements performed on a C14 Nb(Fe$_{89.4}$Al$_{10.6}$)$_2$ compound and provide evidences that its AF structure is not a simple one.

## 2. Experimental

### 2.1. Sample

The Nb-Fe-Al alloy was fabricated from Fe (99.95 wt.%), Nb (99.9 wt.%) and Al (99.999 wt.%) by levitation melting followed by subsequent casting into a pre-heated Cu crucible at 1200°C. Melting, casting and cooling were performed under argon atmosphere. Impurity contents of C, N and O non-metal elements in the alloy were determined to be less than 100 ppm. The phase identification and characterization of the sample obtained was done by XRD patterns. The chemical composition of the final product was determined by EPMA measurements with a Joel JXA-8100 instrument and yielded 57.3 at.% for Fe, 10.6 at.% for Al and 32.1 at.% for Nb. More details on the sample preparation can be found in Ref. 18.

### 2.2. Measurements

ZFC and FC DC magnetization (M) measurements at various applied magnetic fields ($H$ =50 -1250 Oe) in the temperature interval 20 K < $T$ < 180 K, have been performed by a commercial (Quantum Design) superconducting quantum interference device (SQUID) magnetometer with sample mounted in gel-cap. Prior to recording the zero-field-cooled (ZFC) curves, the SQUID magnetometer was always adjusted to be in a "real" $H$ = 0 state. The temperature dependence of the FC branches were taken via warming the sample. Examples of the recorded M(T) curves are displayed in Fig. 1.

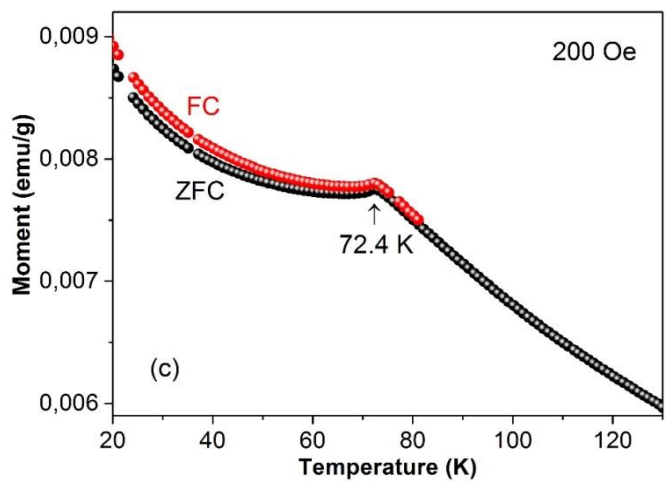

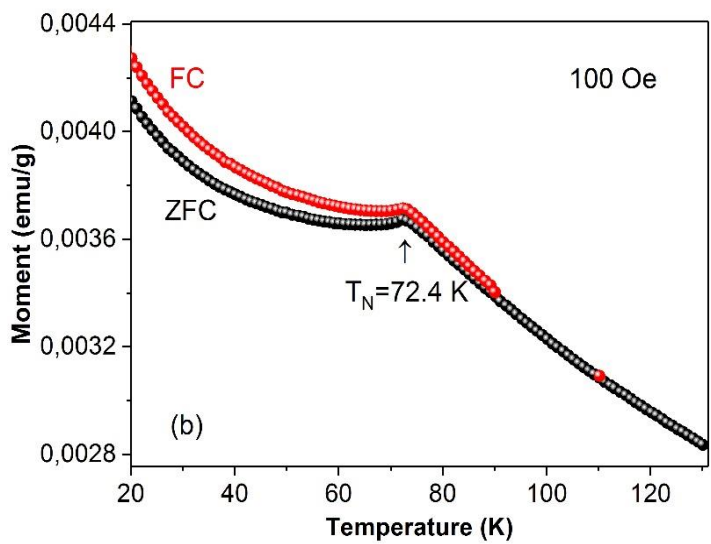

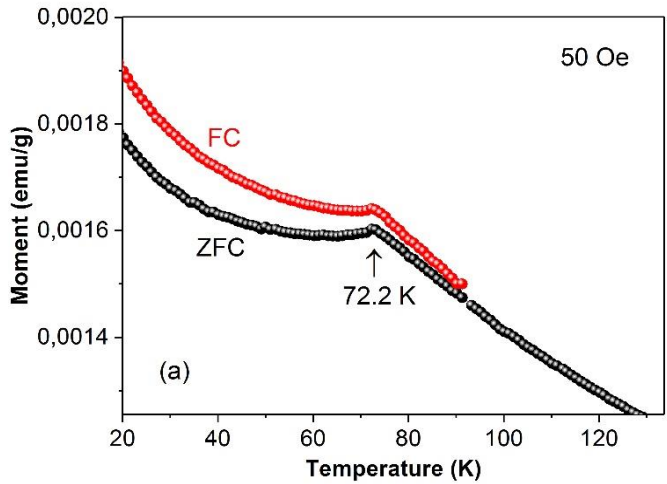

Fig. 1. Zero-field-cooled (ZFC) and field-cooled (FC) magnetization curves vs. temperature under various applied fields shown.

## 3. Results and Discussion

All $M_{ZFC}$ and $M_{FC}$ branches (Fig. 1) exhibit a peak around $T_N$=72.3(2) K clearly indicating an AF magnetic structure. The applied fields do not affect the $T_N$ position as demonstrated in Fig. 2. It is evident from Fig. 1 that the ZFC and FC curves coincide with each other only at sufficiently high temperatures. On lowering temperature the two curves depart from each other at $T_{ir}$, i.e. an irreversibility occurs. The existence of such effect is one of the typical characteristic features for spin-glasses. In any case, this irreversibility indicates that the magnetic state of Nb(Fe$_{89.4}$Al$_{10.6}$)$_2$ is more complex than the simple AF structure reported for Nb(Fe$_{1-x}$Al$_x$)$_2$ with x $\geq$ 0.05 [3]. (The sample in Ref. 9 was ferromagnetic).

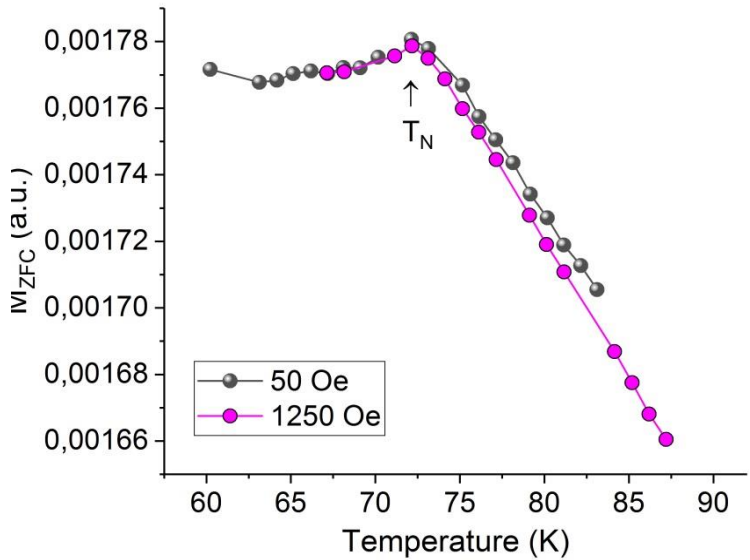

Fig. 2. Normalized ZFC(T) measured in $H$=50 and 1250 Oe. The arrow indicates the constant $T_N$ value.

At high temperatures the curves exhibits typical paramagnetic (PM) shapes and adhere closely to the Curie-Weiss (CW) law: $\chi(T) = \chi_0 + C/(T-\Theta)$, where $\chi(T)=M/H(T)$ and $\chi_0$ is the temperature independent part of the susceptibility $\chi$, $C$ is the Curie constant, and $\Theta$ is the Curie-Weiss (CW) temperature (Fig. 3). The antiferromagnetic character of the spin interactions for $T \geq T_N$ is evidenced by a negative value, $\Theta$, equal to -120 and to -87K for $H$= 50 and 1250 Oe, respectively.

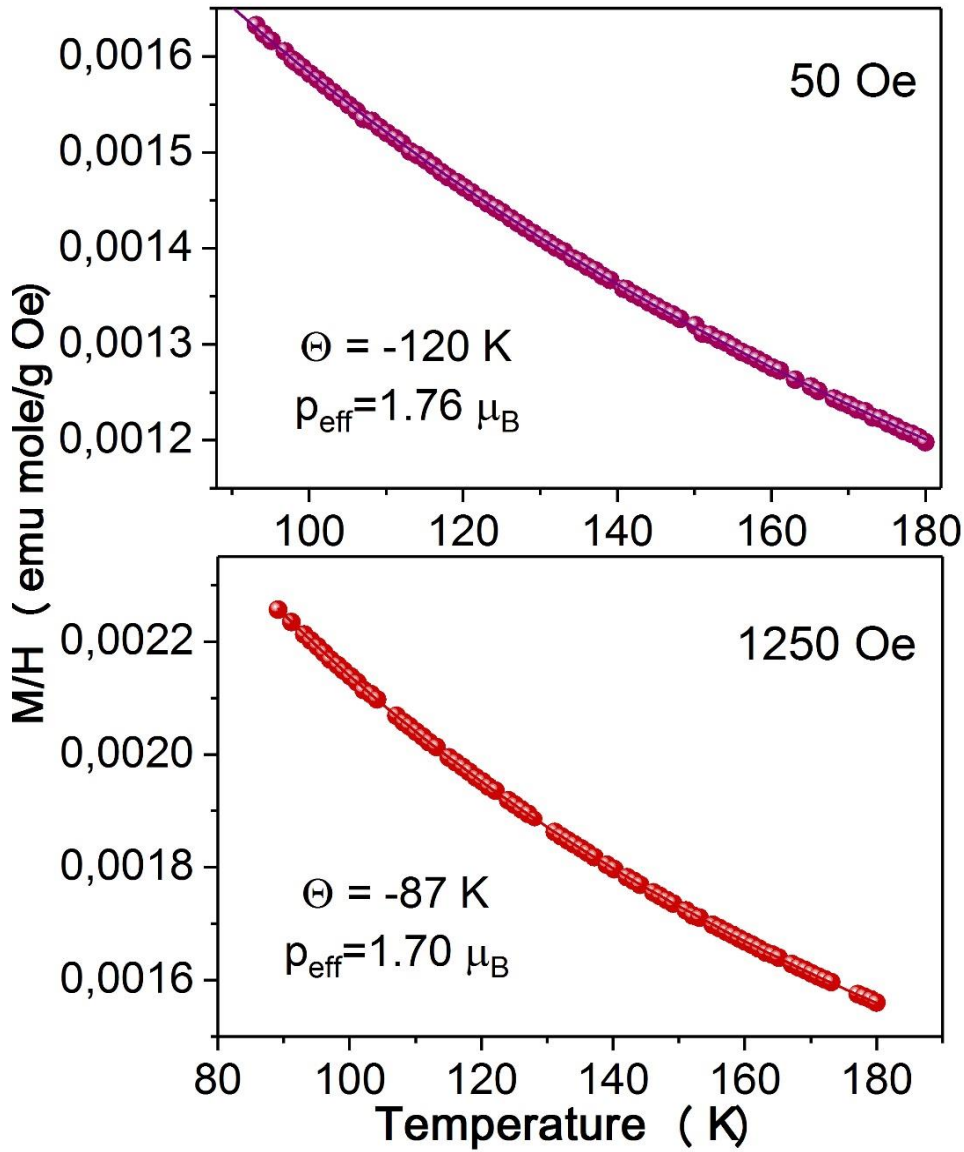

Fig. 3 PM M(T) curves for $T \geq T_N$ measured in the applied magnetic field of 50 Oe and 1250 Oe. The lines stand for the best-fit to the Curie-Weiss law.

From the C values obtained, the PM effective moment given by $p_{eff}=(8C)^{1/2}$ are deduced. The average $p_{eff}$ obtained is 1,74(2) $\mu_B$/ Fe, a value which corresponds to a low spin effective state $2[S(S+1)]^{1/2}\mu_B$ of $Fe^{3+}$. Within the uncertainties this value fits well with 1,83 µ$_B$ obtained for Nb$_{0.975}$Fe$_{2.025}$ in Ref 9. That implies that the low-spin state of Fe$^{3+}$ is not affected by the partial substitution of Fe for Al, although the effective interactions changed upon Al-doping from ferromagnetic to antiferromagnetic ones.

The two Θ values measured for the two extreme values of *H*-field show that: (1) there is some degree of spin frustration (FD) above $T_N$, and (2) the FD value decreases with *H*. Namely, $FD = abs(\frac{\theta}{T_N}) \approx 1.7$ at *H*=50 Oe and *FD*≈1.2 at *H*=1250 Oe. Regarding the irreversibility, it is readily observed in Fig. 1 that it decreases with *H* viz. the difference between the M$_{FC}$ and M$_{ZFC}$ curves becomes smaller. To visualize this effect we calculated $\Delta M = (M_{FC} - M_{ZFC)})/M_{FC}$ for three values of *H* and plotted them in Fig. 4. Noteworthy, for *H*=1250 Oe *ΔM*=0. In addition, Fig.4 shows that the irreversibility temperature, $T_{ir}$, significantly decrease with *H*. This behavior is in line, at least qualitatively, with the mean-field theory prediction, supporting thereby our supposition that an AF-spin-glass like state exists in the temperature range $T_N \leq T \leq T_{ir}$. The increase of both M$_{FC}$ and M$_{ZFC}$ curves for $T < T_N$ i.e. indicates that the magnetic ground state of the investigated compound is a mixture of AF and FM states. However, more systematic studies, including AC susceptibility measurements are needed to better elucidate a full magnetic *H-T* phase diagram.

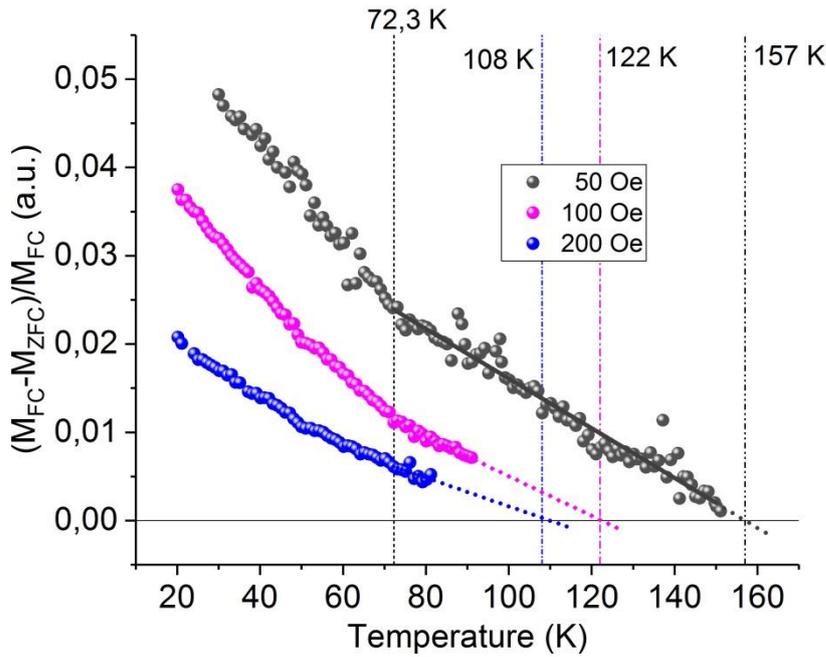

Fig. 4 Relative difference between the $M_{FC}$ and $M_{ZFC}$ divided by $M_{FC}$ curves versus temperature for three different values of the applied magnetic field. The vertical line at the lowest temperature refers to the Néel temperature, while the other three lines indicate the irreversibility temperatures ($T_{ir}$) for the three fields. $T_{ir}$-values were obtained by linear extrapolation to zero the data for $T \geq T_N$.

## 4. Conclusions

Results obtained in the present study permit drawing the following conclusions:

1. The antiferromagnetic structure of the C14 $Nb(Fe_{89.4}Al_{10.6})_2$ Laves phase compound is not a simple one.

2. It orders antiferromagnetically at $T_N=72.3(2)$K but exhibits irreversibility at $T_{ir} > T_N$.

3. The irreversibility temperature ($T_{ir}$) as well as the irreversibility itself significantly decrease with the increase of the applied magnetic field (*H*).

4. Spins in the temperature range above $T_N < T < T_{ir}$ are frustrated and the degree of frustration decreases with $H$.

5. The magnetic ground state of the sample seems to be a mixture of the FM and AF states.

6. The AFSG→AF→FM+AF transitions are the likely magnetic phase transitions scenario in the studied sample.

7. In the paramagnetic range, the PM effective moment ($p_{eff}$) is similar to the corresponding value obtained for $Nb_{0.975}Fe_{2.025}$. The low $p_{eff}$ values obtained correspond to $Fe^{3+}$ in the low spin state

**Acknowledgements**

This study was financed by the Faculty of Physics and Applied Computer Science AGH UST statutory tasks within subsidy of Ministry of Science and Higher Education in Warsaw. The investigated sample was courtesy of Dr. F. Stein.